\PassOptionsToPackage{pagewise}{lineno}
\PassOptionsToPackage{dvipsnames}{xcolor}
\documentclass[submission,Phys]{SciPost}

\usepackage{float}

\usepackage{imakeidx}
\makeindex[columns=2, title=Alphabetical Index, intoc]
\usepackage{lineno}
\usepackage{amsmath,amssymb}
\usepackage{dsfont} 
\usepackage{nicefrac}
\usepackage{xcolor}
\usepackage{braket}
\usepackage{enumitem}
\usepackage{mathtools}
\usepackage{bbold}
\usepackage{physics}

\usepackage{multirow}
\usepackage{caption}
\usepackage{makecell}
\usepackage{subcaption}

\usepackage{graphicx}
\usepackage{tikz-feynman}
\usetikzlibrary{calc}
\tikzfeynmanset{compat=1.1.0,warn luatex=false}
\usepackage{mathdots}
\usepackage{textgreek}

\usepackage{rotating}
\usepackage{cancel}
\usepackage{slashed}
\usepackage{soul}

\usepackage{longtable}
\usepackage{cleveref} 

\definecolor{darkbrown}{RGB}{150,75,10}
\definecolor{dmtgreen}{RGB}{0,120,80}

\usepackage{dowith}

\newcommand{\UpperCase}[1]{
  \expandafter\newcommand\csname bb#1\endcsname{{\mathbb{#1}}}
  \expandafter\newcommand\csname cal#1\endcsname{{\mathcal{#1}}}   
  \expandafter\newcommand\csname rm#1\endcsname{{\mathrm{#1}}}
  \expandafter\newcommand\csname bf#1\endcsname{{\mathbf{#1}}}
  \expandafter\newcommand\csname bold#1\endcsname{{\boldsymbol{#1}}}
  \expandafter\newcommand\csname hat#1\endcsname{\hat{#1}}
  \expandafter\newcommand\csname tilde#1\endcsname{\widetilde{#1}}
  \expandafter\newcommand\csname bar#1\endcsname{\overline{#1}}
  \expandafter\newcommand\csname frak#1\endcsname{\mathfrak{#1}}
  }
\DoWith\UpperCase ABCDEFGHILJKMNOPQRSTUVWXYZ\StopDoing

\newcommand{\LowerCase}[1]{
  \expandafter\newcommand\csname rm#1\endcsname{{\mathrm{#1}}} 
  \expandafter\newcommand\csname bf#1\endcsname{{\mathbf{#1}}} 
  \expandafter\newcommand\csname bold#1\endcsname{{\boldsymbol{#1}}}
  \expandafter\newcommand\csname hat#1\endcsname{\hat{#1}}
  \expandafter\newcommand\csname tilde#1\endcsname{\tilde{#1}}
  \expandafter\newcommand\csname bar#1\endcsname{\bar{#1}}
  \expandafter\newcommand\csname frak#1\endcsname{\mathfrak{#1}}
  \expandafter\newcommand\csname vec#1\endcsname{\vec{#1}}
  }
\DoWith\LowerCase abcdefghijklmnopqrstuvwxyz\StopDoing

\newcommand{\eq}{Eq.~}

\newcommand{\dx}{\rmd x}

\newcommand{\Fp}{\mathfrak{m}}
\newcommand{\Sp}{\mathfrak{n}}
\usepackage[normalem]{ulem}

\usepackage{dashrule}

\usepackage{array}
\usepackage{xcolor}
\usepackage{colortbl}
\usepackage{graphicx}
\usepackage{booktabs}

\definecolor{AccentFill}{RGB}{240,240,240}
\definecolor{SectionFill}{RGB}{247,247,247}

\newcommand{\HeaderFont}{\normalfont\scshape}
\newcommand{\SelectionFont}{\normalfont\itshape}

\newlength{\LabelWidth}
\newlength{\MetricWidth}
\newlength{\FirstCaseWidth}
\newlength{\CaseWidth}
\newlength{\ThirdCaseWidth}
\newlength{\HeaderHeight}
\newlength{\OrderHeight}
\newlength{\BodyHeight}
\newlength{\MetricSpacing}
\newlength{\SeparatorWidth}
\newlength{\PanelSeparatorTopInset}
\newlength{\ProcessDividerThickness}
\newlength{\ProcessDividerGap}
\newcolumntype{C}[1]{w{c}{#1}}

\newsavebox{\VBFTableBox}
\newlength{\HeaderBlockHeight}

\newcommand{\ResultCell}[3]{%
    \parbox[c][\BodyHeight][c]{\CaseWidth}{%
        \centering
        \setlength{\baselineskip}{\MetricSpacing}%
        \strut #1\par
        \strut #2\par
        \strut #3\par
    }%
}

\newcommand{\MetricLabels}{%
    \parbox[c][\BodyHeight][c]{%
        \dimexpr\MetricWidth-4mm\relax
    }{%
        \raggedright
        \setlength{\baselineskip}{\MetricSpacing}%
        \strut\mbox{$\sigma \pm \delta\sigma$ (fb)}\par
        \strut\mbox{CPU core-hours}\par
        \strut Integr.\ difficulty $\calE/\calE_0$ \par
    }%
}

\newcommand{\SelectionLabel}[1]{%
    \rotatebox[origin=c]{90}{\SelectionFont #1}%
}

\newcommand{\CaseHeading}[1]{%
    \parbox[c][\HeaderHeight][c]{\CaseWidth}{%
        \centering
        \HeaderFont #1%
    }%
}

\newcommand{\OrderHeading}[1]{%
    \parbox[c][\OrderHeight][c]{%
        \dimexpr
        \FirstCaseWidth+2\CaseWidth+\ThirdCaseWidth
        \relax
    }{%
        \centering
        \bfseries #1%
    }%
}

\newcommand{\VBFTableStretch}{0.75}

\newcommand{\VBFComparisonPanel}[4]{%
\begingroup
\setlength{\tabcolsep}{0pt}%
\setlength{\aboverulesep}{0.6mm}%
\setlength{\belowrulesep}{0.6mm}%
\setlength{\cmidrulesep}{0pt}%

\setlength{\HeaderHeight}{\VBFTableStretch\HeaderHeight}%
\setlength{\OrderHeight}{\VBFTableStretch\OrderHeight}%
\setlength{\BodyHeight}{\VBFTableStretch\BodyHeight}%
\setlength{\MetricSpacing}{\VBFTableStretch\MetricSpacing}%
\setlength{\aboverulesep}{\VBFTableStretch\aboverulesep}%
\setlength{\belowrulesep}{\VBFTableStretch\belowrulesep}%
\setlength{\PanelSeparatorTopInset}{#4}%

\ifdim\HeaderHeight<10mm
    \setlength{\HeaderHeight}{10mm}%
\fi

\renewcommand{\OrderHeading}[1]{%
    \parbox[c][\OrderHeight][c]{%
        \dimexpr\FirstCaseWidth+2\CaseWidth+\ThirdCaseWidth\relax
    }{\centering\bfseries ##1}%
}%

\setlength{\HeaderBlockHeight}{%
    \dimexpr
    \HeaderHeight+\OrderHeight
    +2\aboverulesep+2\belowrulesep+1.6pt
    \relax
}%

\normalfont\fontsize{10}{12}\selectfont
\sbox{\VBFTableBox}{%
\begin{tabular}{@{}
    C{\LabelWidth}
    C{\MetricWidth}
    C{\FirstCaseWidth}
    C{\CaseWidth}
    C{\CaseWidth}
    C{\ThirdCaseWidth}
@{}}

    #3
    \multicolumn{2}{c}{}
    & \CaseHeading{\shortstack{Humans\\{\fontsize{9.5}{10}\selectfont(old paramet.)}}}
    & \CaseHeading{\shortstack{Humans\\{\fontsize{9.5}{10}\selectfont(new paramet.)}}}
    & \CaseHeading{\shortstack{LLM\\{\fontsize{9.5}{10}\selectfont(guided)}}}
    & \CaseHeading{\shortstack{LLM\\{\fontsize{9.5}{10}\selectfont(autonomous)}}}
    \\ \cmidrule[0.4pt]{3-6}

    #2

\end{tabular}%
}

\usebox{\VBFTableBox}%

\llap{%
    \makebox[\wd\VBFTableBox][l]{%
        \raisebox{%
            \dimexpr
            \ht\VBFTableBox-\aboverulesep
            -0.5\HeaderBlockHeight
            \relax
        }{%
            \makebox[%
                \dimexpr\LabelWidth+\MetricWidth\relax
            ][c]{%
                \raisebox{-0.5\height}{\large\bfseries #1}%
            }%
        }%
    }%
}%

\llap{%
    \makebox[\wd\VBFTableBox][l]{%
        \hspace*{\dimexpr\LabelWidth+\MetricWidth\relax}%
        \raisebox{-\dp\VBFTableBox}{%
            \rule{\SeparatorWidth}{%
                \dimexpr
                \ht\VBFTableBox+\dp\VBFTableBox
                -\PanelSeparatorTopInset
                \relax
            }%
        }%
    }%
}%

\endgroup
}

\newcommand{\VBFComparisonRows}{%

    \multicolumn{2}{c}{}
    & \multicolumn{4}{c}{
        \cellcolor{SectionFill}\OrderHeading{LO}
    }
    \\ \midrule[0.4pt]

    \SelectionLabel{Inclusive}
    & \cellcolor{AccentFill}\MetricLabels
    & \ResultCell{$4091.2 \pm 1.0$}{0.24}{1.5} 
    & \ResultCell{$4092.4 \pm 1.0$}{0.061}{0.37} 
    & \ResultCell{$4092.8 \pm 1.0$}{0.033}{0.20} 
    & \ResultCell{$4093.0 \pm 1.0$}{0.071}{0.43} 
    \\
    \arrayrulecolor{black!55}
    \midrule[0.07pt]
    \arrayrulecolor{black}

    \SelectionLabel{Fiducial}
    & \cellcolor{AccentFill}\MetricLabels
    & \ResultCell{$2776.4 \pm 1.0$}{0.13}{1.8} 
    & \ResultCell{$2774.9 \pm 1.0$}{0.026}{0.34} 
    & \ResultCell{$2775.0 \pm 1.0$}{0.026}{0.34} 
    & \ResultCell{$2776.1 \pm 1.0$}{0.041}{0.53} 
    \\ \midrule[0.4pt]

    \multicolumn{2}{c}{}
    & \multicolumn{4}{c}{
        \cellcolor{SectionFill}\OrderHeading{NLO}
    }
    \\ \midrule[0.4pt]

    \SelectionLabel{Inclusive}
    & \cellcolor{AccentFill}\MetricLabels
    & \ResultCell{$-54.3 \pm 1.0$}{32}{197} 
    & \ResultCell{$-51.7 \pm 1.0$}{4.9}{31} 
    & \ResultCell{$-53.1 \pm 1.0$}{0.67}{4.0} 
    & \ResultCell{$-52.2 \pm 1.0$}{2.9}{18} 
    \\
    \arrayrulecolor{black!55}
    \midrule[0.07pt]
    \arrayrulecolor{black}

    \SelectionLabel{Fiducial}
    & \cellcolor{AccentFill}\MetricLabels
    & \ResultCell{$-95.8 \pm 1.0$}{13}{176} 
    & \ResultCell{$-95.4 \pm 1.0$}{2.4}{31} 
    & \ResultCell{$-94.6 \pm 1.0$}{0.80}{10} 
    & \ResultCell{$-95.3 \pm 1.0$}{2.4}{32} 
    \\ \midrule[0.4pt]

    \multicolumn{2}{c}{}
    & \multicolumn{4}{c}{
        \cellcolor{SectionFill}\OrderHeading{NNLO}
    }
    \\ \midrule[0.4pt]

    \SelectionLabel{Inclusive}
    & \cellcolor{AccentFill}\MetricLabels
    & \ResultCell{$-41.6 \pm 1.0$}{1208}{7443} 
    & \ResultCell{$-39.9 \pm 1.0$}{58}{349} 
    & \ResultCell{$-40.0 \pm 1.0$}{35}{208} 
    & \ResultCell{$-40.6 \pm 1.0$}{70}{406} 
    \\
    \arrayrulecolor{black!55}
    \midrule[0.07pt]
    \arrayrulecolor{black}

    \SelectionLabel{Fiducial}
    & \cellcolor{AccentFill}\MetricLabels
    & \ResultCell{$-112.2 \pm 1.0$}{369}{4915} 
    & \ResultCell{$-109.6 \pm 1.0$}{43}{572} 
    & \ResultCell{$-111.4 \pm 1.0$}{33}{427} 
    & \ResultCell{$-111.3 \pm 1.0$}{90}{1160} 
    \\ \bottomrule[0.8pt]
}

\newcommand{\VBFJetComparisonRows}{%

    \multicolumn{2}{c}{}
    & \multicolumn{4}{c}{
        \cellcolor{SectionFill}\OrderHeading{LO}
    }
    \\ \midrule[0.4pt]

    \SelectionLabel{Fiducial}
    & \cellcolor{AccentFill}\MetricLabels
    & \ResultCell{--}{--}{--} 
    & \ResultCell{$571.3 \pm 1.0$}{0.055}{16} 
    & \ResultCell{$572.5 \pm 1.0$}{0.028}{8.5} 
    & \ResultCell{$573.8 \pm 1.0$}{0.018}{5.0} 
    \\ \midrule[0.4pt]

    \multicolumn{2}{c}{}
    & \multicolumn{4}{c}{
        \cellcolor{SectionFill}\OrderHeading{NLO}
    }
    \\ \midrule[0.4pt]

    \SelectionLabel{Fiducial}
    & \cellcolor{AccentFill}\MetricLabels
    & \ResultCell{--}{--}{--} 
    & \ResultCell{$37.0 \pm 1.0$}{10}{3073} 
    & \ResultCell{$36.6 \pm 1.0$}{1.1}{336} 
    & \ResultCell{$37.1 \pm 1.0$}{20}{5800} 
    \\ \midrule[0.4pt]

    \multicolumn{2}{c}{}
    & \multicolumn{4}{c}{
        \cellcolor{SectionFill}\OrderHeading{NNLO}
    }
    \\ \midrule[0.4pt]

    \SelectionLabel{Fiducial}
    & \cellcolor{AccentFill}\MetricLabels
    & \ResultCell{--}{--}{--} 
    & \ResultCell{$-32.3 \pm 1.0$}{1348}{$4.1 \times 10^{5}$} 
    & \ResultCell{$-31.4 \pm 1.0$}{1128}{$3.4 \times 10^{5}$} 
    & \ResultCell{$-30.9 \pm 1.0$}{1978}{$6.2 \times 10^{5}$} 
    \\ \bottomrule[0.8pt]
}

\newcommand{\VBFComparisonTable}{%
    \VBFComparisonPanel{WBF}{\VBFComparisonRows}{\cmidrule[0.8pt]{3-6}}{\aboverulesep}%
}

\newcommand{\VBFJetComparisonTable}{%
    \VBFComparisonPanel{WBF+$j$}{\VBFJetComparisonRows}{\cmidrule[0.8pt]{3-6}}{\aboverulesep}%
}

\newcommand{\VBFImplementationCell}[1]{%
    \parbox[c]{\dimexpr\CaseWidth-8mm\relax}{%
        \footnotesize
        \vspace*{0.45mm}%
        \strut #1\strut\par
        \vspace*{0.45mm}%
    }%
}

\newcommand{\VBFImplementationRow}[3]{%
    \cellcolor{AccentFill}%
    \parbox[c][5.5mm][c]{\dimexpr\MetricWidth-6mm\relax}{%
        \footnotesize\bfseries\strut #1\strut
    }%
    & \VBFImplementationCell{#2} & \VBFImplementationCell{#3} \\
}

\newcommand{\VBFImplementationWBFRows}{%
    \VBFImplementationRow{Inputs}
        {Refs~\cite{Devoto:2023rpv,Devoto:2025kin,Devoto:2025jql}; optimized amplitudes; unpublished analytical inputs (including the limits below); WBF code of Ref.~\cite{Asteriadis:2021gpd} for benchmarking.}
        {Refs~\cite{Devoto:2023rpv,Devoto:2025kin,Devoto:2025jql}; optimized amplitudes, inclusive Higgs NNLO code and the preceding Higgs+$j$ implementation.}
    \arrayrulecolor{black!55}\midrule[0.07pt]
    \VBFImplementationRow{Additional limits}
        {General analytical expressions supplied by us.}
        {Derived by the agents.}
    \midrule[0.07pt]
    \VBFImplementationRow{Execution}
        {Local development on a laptop; jobs submitted to a cluster by a human expert, with their output provided to the agents.}
        {Direct cluster access for code development, execution and analysis of numerical output.}
    \midrule[0.07pt]
    \VBFImplementationRow{Reference checks}
        {Comparison with WBF results from Ref.~\cite{Asteriadis:2021gpd}, with numerical feedback from us.}
        {Inclusive photon-mediated DIS against \texttt{Dis\-order}~\cite{Karlberg:2024hnl} and \texttt{HOPPET}~\cite{Salam:2008qg}, one-jet DIS through NNLO against \texttt{NNLO\-JET}~\cite{NNLOJET:2025rno}, in\-clu\-sive WBF through NNLO against \texttt{proVBFH-\allowbreak inclu\-sive}~\cite{Dreyer:2016oyx}, and inclusive and fi\-du\-cial WBF at LO and NLO against \texttt{MCFM}~\cite{Campbell:2019MCFM} and \texttt{VBFNLO}~\cite{Arnold:2008rz}.}
    \midrule[0.07pt]\arrayrulecolor{black}
    \VBFImplementationRow{Human involvement}
        {Expert guidance on development, verifying conventions and optimizations, and investigating numerical instabilities and discrepancies.}
        {Largely unsupervised development; human input comprised task definition, provision of inputs and requests for further optimization.}
}

\newcommand{\VBFImplementationWBFJetRows}{%
    \VBFImplementationRow{Starting code}
        {The corresponding WBF code and inputs, supplemented with WBF+$j$ amplitudes.}
        {The corresponding WBF code and inputs, supplemented with WBF+$j$ amplitudes.}
    \arrayrulecolor{black!55}\midrule[0.07pt]
    \VBFImplementationRow{Additional limits}{General analytical expressions supplied by us.}{Derived by the agents.}
    \midrule[0.07pt]
    \VBFImplementationRow{Execution}
        {Same as for WBF.}
        {Same as for WBF.}
    \midrule[0.07pt]
    \VBFImplementationRow{Reference checks}
        {No external benchmark.}
        {WBF+$j$ at LO against \texttt{VBFNLO}~\cite{Arnold:2008rz}.}
    \midrule[0.07pt]\arrayrulecolor{black}
    \VBFImplementationRow{Human involvement}
        {Same as for WBF.}
        {Same as for WBF.}
    \bottomrule[0.8pt]
}

\newcommand{\VBFImplementationPanel}[4]{%
    \sbox{\VBFTableBox}{%
        \begin{tabular}{@{}
            C{\MetricWidth}
            C{\CaseWidth}
            C{\CaseWidth}
        @{}}
            #3
            \parbox[c][\HeaderHeight][c]{\MetricWidth}{\centering\large\bfseries #1}
            & \CaseHeading{\shortstack{LLM\\{\fontsize{9.5}{10}\selectfont(guided)}}}
            & \CaseHeading{\shortstack{LLM\\{\fontsize{9.5}{10}\selectfont(autonomous)}}}
            \\ \midrule[0.4pt]
            #2
        \end{tabular}%
    }%
    \usebox{\VBFTableBox}%
    \llap{%
        \makebox[\wd\VBFTableBox][l]{%
            \hspace*{\MetricWidth}%
            \raisebox{-\dp\VBFTableBox}{%
                \rule{\SeparatorWidth}{%
                    \dimexpr\ht\VBFTableBox+\dp\VBFTableBox-#4\relax
                }%
            }%
        }%
    }%
}

\newcommand{\VBFImplementationTable}{%
    \begingroup
    \setlength{\tabcolsep}{0pt}%
    \setlength{\aboverulesep}{0.6mm}%
    \setlength{\aboverulesep}{\VBFTableStretch\aboverulesep}%
    \setlength{\belowrulesep}{\aboverulesep}%
    \setlength{\cmidrulesep}{0pt}%
    \setlength{\HeaderHeight}{10mm}%
    \normalfont\fontsize{10}{12}\selectfont
    \setlength{\CaseWidth}{%
        \dimexpr(\LabelWidth+\MetricWidth+\FirstCaseWidth+2\CaseWidth+\ThirdCaseWidth)/2\relax
    }%
    \settowidth{\MetricWidth}{\footnotesize\bfseries Human involvement}%
    \addtolength{\MetricWidth}{6mm}%
    \addtolength{\CaseWidth}{-\dimexpr\MetricWidth/2\relax}%
    \vbox{%
        \offinterlineskip
        \hbox{\VBFImplementationPanel{WBF}{\VBFImplementationWBFRows}{%
            \cmidrule[0.8pt]{2-3}%
        }{\aboverulesep}}%
        \hbox{\VBFImplementationPanel{WBF+$j$}{\VBFImplementationWBFJetRows}{%
            \specialrule{\ProcessDividerThickness}{0pt}{\ProcessDividerGap}%
            \specialrule{\ProcessDividerThickness}{0pt}{\belowrulesep}%
        }{\dimexpr2\ProcessDividerThickness+\ProcessDividerGap\relax}}%
    }%
    \endgroup
}

\usetikzlibrary{patterns}
\newcommand{\WBFUpperLinePanel}[1]{%
\begin{tikzpicture}[baseline={(0,0)}]
\begin{feynman}
\vertex (iU) at (-2,0.8) {$q$};
\vertex[blob,minimum size=6.05mm,preaction={fill=black!18},pattern=north east lines,pattern color=black] (u) at (0,0.8) {};
\vertex (oU) at (2,0.8) {$q'$};
\vertex (iL) at (-2,-0.8) {$Q$};
\vertex (l) at (0,-0.8);
\vertex (oL) at (2,-0.8) {$Q'$};
\vertex (v) at (0,0);
\vertex (h) at (2,0) {$H$};
\diagram*{
(iU) -- [plain] (u) -- [plain] (oU),
(iL) -- [plain] (l) -- [plain] (oL),
(u) -- [photon, edge label'={$V$}] (v),
(l) -- [photon, edge label={$V$}] (v),
(v) -- [scalar] (h),
};
\foreach \current in {u,l}{
  \foreach \arrowX in {-1,1}{
    \path[/tikzfeynman/with arrow=0.5]
      ([xshift=\arrowX cm]\current.center) ++(-1mm,0) -- ++(2mm,0);
  }
}
\ifnum#1=1
\vertex (g) at (1.65,1.82) {$g$};
\diagram*{(g) -- [gluon] (u)};
\fi
\end{feynman}
\end{tikzpicture}}
\newcommand{\WBFBornFigure}{\scalebox{0.88}{\WBFUpperLinePanel{0}}\hspace{16mm}\scalebox{0.88}{\WBFUpperLinePanel{1}}}

\newcommand{\be}{\begin{equation}}
\newcommand{\ee}{\end{equation}}
\makeatletter
\newcommand\ps@indexpagestyle{
 \renewcommand\@oddfoot{\hfill-- \thepage\ --\hfill}
 \renewcommand\@oddhead{}
}

\let\oldsqrt\sqrt
\def\sqrt{\mathpalette\DHLhksqrt}
\def\DHLhksqrt#1#2{%
\setbox0=\hbox{$#1\oldsqrt{#2\,}$}\dimen0=\ht0
\advance\dimen0-0.3\ht0
\setbox2=\hbox{\vrule height\ht0 depth -\dimen0}%
{\box0\lower0.4pt\box2}}

\providecommand{\eprint}[2][]{%
  \href{https://arxiv.org/abs/#2}{arXiv:\nolinkurl{#2}}%
}

\definecolor{brightpink}{rgb}{1.0, 0.08, 0.58} 

\makeatother
\indexsetup{firstpagestyle=indexpagestyle}

\def\SLAC{SLAC National Accelerator Laboratory, Stanford University, Stanford, CA 94039, USA}
\def\KITA{Institute for Theoretical Particle Physics, KIT, Wolfgang-Gaede-Straße 1, 76131, Karlsruhe, Germany}
\def\TIF{Tif Lab, Dipartimento di Fisica, Università di Milano and
INFN, Sezione di Milano, Via Celoria 16, I-20133 Milano, Italy}
\def\CERN{Theoretical Physics Department, CERN, 1211 Geneva 23, Switzerland}

\newcommand{\manuscriptpreprint}{
\begin{flushright} 
SLAC-PUB-261123, TTP26-044, P3H-26-095, TIF-UNIMI-2026-19, CERN-TH-2026-299
\end{flushright}
}

\newcommand{\manuscripttitle}{
From nested soft-collinear subtraction to NNLO QCD predictions with large language models
}

\newcommand{\manuscriptauthors}{
Federica Devoto\textsuperscript{a},
Kirill Melnikov\textsuperscript{b},
Ivan Novikov\textsuperscript{b},
Raoul R{\"o}ntsch\textsuperscript{c},
Chiara Signorile-Signorile\textsuperscript{d},
Davide Maria Tagliabue\textsuperscript{b} and
Matteo Tresoldi\textsuperscript{b}
}

\newcommand{\manuscriptemails}{
\href{mailto:federica@slac.stanford.edu}{federica@slac.stanford.edu},\\[2pt]
\href{mailto:kirill.melnikov@kit.edu}{kirill.melnikov@kit.edu},\\[2pt]
\href{mailto:ivan.novikov@kit.edu}{ivan.novikov@kit.edu},\\[2pt]
\href{mailto:raoul.rontsch@unimi.it}{raoul.rontsch@unimi.it},\\[2pt]
\href{mailto:chiara.signorile-signorile@cern.ch}{chiara.signorile-signorile@cern.ch},\\[2pt]
\href{mailto:davide.tagliabue@kit.edu}{davide.tagliabue@kit.edu},\\[2pt]
\href{mailto:matteo.tresoldi@partner.kit.edu}{matteo.tresoldi@partner.kit.edu}
}

\newcommand{\manuscriptaffiliations}{
\textbf{a} \SLAC \\[2pt]
\textbf{b} \KITA \\[2pt]
\textbf{c} \TIF \\[2pt]
\textbf{d} \CERN
}

\newcommand{\manuscriptabstract}{ 
We investigate whether large language models   can construct reliable and efficient codes for computing NNLO QCD corrections to collider processes, starting from the general formula  derived within the nested soft-collinear subtraction scheme.
We use Higgs production in weak-boson fusion with and without an additional jet to compare numerical implementations by large language models, subject to various degrees of supervision, against a human-written reference code.  
}

\newcommand{\manuscriptkeywords}{QCD corrections, NNLO calculations, infrared subtraction schemes, large language models}

\allowdisplaybreaks 
\hypersetup{hidelinks}
\begin{document}

\manuscriptpreprint

\begin{center}
{\Large\bfseries \manuscripttitle\par}
\end{center}

\begin{center}
\manuscriptauthors
\end{center}

\begin{center}
\small
\manuscriptaffiliations
\end{center}

\begin{center}
\small\sffamily
\manuscriptemails
\end{center}

\section*{Abstract}
{\bfseries \manuscriptabstract\par}

\smallskip
\noindent\textbf{Keywords:} \manuscriptkeywords

\vspace{10pt}
\noindent\rule{\textwidth}{1pt}
\tableofcontents\thispagestyle{fancy}
\noindent\rule{\textwidth}{1pt}
\vspace{10pt}
\raggedbottom 

\section{Introduction}
\label{sec:introduction}

Recently, we presented a general formula for computing next-to-next-to-leading order (NNLO) QCD corrections to arbitrary collider processes with massless partons within the nested soft-collinear subtraction scheme~\cite{Devoto:2025jql},
completing the research program initiated in Refs~\cite{Devoto:2023rpv,Devoto:2025kin,Caola:2017dug,Caola:2019nzf,Caola:2019pfz,Caola:2018pxp,Delto:2019asp,Asteriadis:2021gpd}.
The resulting formula requires only a minimal set of process-specific inputs (such as matrix elements) to be supplied in each case.
As such, it can serve as the foundation of a numerical program
for NNLO QCD computations for many collider processes.

General numerical frameworks of this kind are well established at
next-to-leading order (NLO). For example, Catani--Seymour dipole
subtraction~\cite{Catani:1996jh} was key to the construction of
\texttt{MCFM}~\cite{Campbell:1999ah}, a repository of process-specific
routines that grew to support a few hundred processes at NLO QCD
accuracy, using supplied one-loop and real-emission amplitudes,
phase-space parametrizations and histogram routines. 
A complementary example
is \texttt{MadGraph5\_aMC@NLO}~\cite{Alwall:2014hca}, which combines
FKS subtraction~\cite{Frixione:1995ms,Frixione:1997np} with automated
generation of tree-level and one-loop amplitudes to compute NLO QCD
corrections for user-specified processes.

Following an approach similar to that of \texttt{MCFM}, we have
been developing a general numerical program for NNLO QCD computations based on
the universal NNLO subtraction formula of Ref.~\cite{Devoto:2025jql}.
We have designed and
implemented a modular code architecture, and optimized and
debugged the resulting program.

We have pursued this development
for about a year, using Higgs production in weak-boson fusion
(WBF), with and without an additional jet, as a testing ground for a general code applicable to a wide range of collider processes.
Computing NNLO QCD corrections to WBF-like processes is a challenging task: their leading-order
(LO) final states already contain several particles, the phase
space is complex, and the dominant kinematics is forward, leading to delicate cancellations between different contributions.

While pursuing this research program, we have witnessed rapid progress in large language models (LLMs) and their increasing use in scientific computing. With the general formula of Ref.~\cite{Devoto:2025jql} making the subtraction structure explicit, it is natural to ask whether these models can carry out the implementation work described above, optimizing and standardizing human effort.
Specifically, we would like to explore the following questions:
\begin{enumerate}[label=\roman*)]
    \item Can coding agents use the general formula of
    Ref.~\cite{Devoto:2025jql}
    as a starting point for constructing
    working codes to compute NNLO QCD corrections for a
    specific process?
    
    \item What validation and expert guidance are needed to establish
    the reliability of the resulting calculations?
    
    \item How do the resulting codes compare with the human
    implementation in terms of numerical efficiency?
    
    \item What can we learn
    from the agents' choices?
\end{enumerate}

Our attempts to answer these questions are the subject of this paper, which is organized as follows. In Section~\ref{sec:problem}, we briefly explain how perturbative QCD computations are performed, providing the physics background for the subsequent discussion. In Section~\ref{sec:development}, we describe the setup of our experiment with coding agents. In Section~\ref{sec:development_lessons}, we discuss
notable insights from LLM-assisted implementations of the NNLO QCD subtraction method. In Section~\ref{sec_Results}, we compare the numerical results and integration performance of the human and LLM-generated codes for WBF and WBF+$j$ processes. We conclude in Section~\ref{sec_Conclusions}. Some technical
details are given in Appendix~\ref{app:numerical_setup}.

\section{The underlying problem}
\label{sec:problem}

We briefly describe the computation of differential cross sections for collider
processes in perturbative QCD, emphasizing the aspects relevant to the present study.
This is a well-researched problem, but it becomes increasingly complex
as one moves to higher perturbative orders. Currently, leading-order (LO) and NLO computations in QCD are fully automated \cite{Catani:1996jh,Campbell:1999ah,Campbell:2019MCFM,Alwall:2014hca,Ossola:2007ax,Bevilacqua:2011xh,Czakon:2009ss} and there has been tremendous progress in NNLO QCD computations (see, e.g., Refs~\cite{Chen:2014gva,Boughezal:2015dra,Caola:2015wna,Chen:2016zka,Campbell:2019gmd,Cacciari:2015jma,Cruz-Martinez:2018rod,Gauld:2021ule,Catani:2022mfv,Chawdhry:2019bji,Chawdhry:2021hkp,Czakon:2020coa,Gauld:2023zlv,Currie:2017eqf,Chen:2022tpk,Badger:2023mgf,Czakon:2021mjy,Czakon:2015owf,Catani:2019hip,Buonocore:2023ljm,Brucherseifer:2014ama,Berger:2016oht,Campbell:2020fhf,Bronnum-Hansen:2022tmr,Buonocore:2022pqq,Alvarez:2023fhi,Devoto:2024nhl,Buonocore:2025fqs,Bonino:2025qta} for a selection of phenomenological papers employing different theoretical methods). Furthermore, N$^3$LO QCD predictions are now available for a
growing range of collider processes~\cite{Anastasiou:2015vya,Mistlberger:2018etf,Dreyer:2016oyx,Currie:2018fgr,Duhr:2019kwi,Duhr:2020seh,Duhr:2020sdp,Chen:2021isd,Chen:2022cgv,Chen:2022lwc,Baglio:2022wzu,Chen:2025kez,Chen:2026zmi,Czakon:2026zhi,Gehrmann-DeRidder:2026edl}.

As an example, we consider NNLO QCD corrections to a process $pp \to \rmX$, where $\rmX$ contains $N$ massless partons at tree level that
are identified with final-state jets.\footnote{The process might also involve an arbitrary number of colorless final-state particles; this does not affect the discussion in this section.}
To obtain the NNLO contribution to this process, one computes two-loop virtual corrections to the partonic process
$ij \to \rmX$, a one-loop correction to the partonic
process $ij \to \rmX+\Fp$ and accounts for the double-real-emission process $ij \to \rmX+\Fp+\Sp$, 
where $\Fp, \Sp$ are colored particles.
Thus, if one is interested in computing an infrared-safe observable $\calO$ in such a process,
the NNLO contribution
can be written schematically as
\begin{equation}
\begin{split}
    \rmd\sigma_{pp \to \rmX}^{\text{\tiny NNLO}}(\calO)
    = & \sum_{ij}
    \int \dx_1 \dx_2 \, f_i(x_1) f_j (x_2)
    \bigg[\int
    \rmd\Phi_N \, \rmX^{\rm VV}_{ij} \, {\calO}(\Phi_N) \\
    & + \int \rmd\Phi_{N+1} \, \rmX^{\rm RV}_{ij} \, \calO(\Phi_{N+1})
    + \int \rmd\Phi_{N+2} \, \rmX^{\rm RR}_{ij} \, \calO(\Phi_{N+2})
    \bigg] \,.
\end{split}
\label{eq:bare}
\end{equation}
Here, $f_i(x_1)$ and $f_j(x_2)$ are parton distribution functions, $\rmX^{\rm VV}_{ij}$, $\rmX^{\rm RV}_{ij}$ and $\rmX^{\rm RR}_{ij}$ denote the double-virtual, real-virtual and double-real matrix-element contributions, respectively, including the relevant interferences and squared amplitudes, and
$\rmd\Phi_{N+k}$, with $k=0,1,2$, describes the phase space for $N+k$ final-state partons and any accompanying colorless particles.

One of the main difficulties in computing NNLO QCD corrections is that the three individual contributions on the right-hand side of \eq\eqref{eq:bare} are infrared divergent. These divergences
manifest themselves differently.
If dimensional regularization is used, the virtual amplitudes contain explicit infrared poles in $\epsilon = (4-d)/2$. The
double-real terms contain non-integrable singularities at phase-space boundaries and hence require an infrared regulator.
The real-virtual contribution contains both explicit and phase-space-boundary singularities.
These singularities prevent a direct Monte Carlo evaluation of the individual
contributions in four dimensions.

This problem is solved by the so-called subtraction methods \cite{Catani:1996vz,Frixione:1995ms,Frixione:1997np,Gehrmann-DeRidder:2005btv,Somogyi:2006da,Somogyi:2006db,Magnea:2018hab,Bertolotti:2022aih,Caola:2017dug,Czakon:2010td,Boughezal:2011jf} which reorganize \eq\eqref{eq:bare} by introducing local counterterms that
reproduce the singular behavior of the real-virtual and double-real
contributions in unresolved limits, rendering the corresponding
subtracted contributions integrable over phase space.
The subtraction terms must be added back
after integration over the unresolved phase space, leading to additional functions which also contain poles in the regulator $\epsilon$. Upon combining all such functions, and including PDF renormalization, one verifies the cancellation of infrared poles and obtains a finite result as $\epsilon \to 0$.

A serious challenge at NNLO is the overlapping nature of different singular limits.
In the context of nested soft-collinear subtraction, to disentangle overlapping singularities we introduce various partition functions~\cite{Frixione:1995ms,Frixione:1997np},
angular sectors and a parametrization
of the double-unresolved phase space~\cite{Czakon:2010td, Czakon:2011ve}
which avoids the double counting of the singular
limits.
Ref.~\cite{Devoto:2025jql} gives this construction in the form of a general, regulator-independent formula for computing NNLO QCD corrections to arbitrary collider processes with massless partons.
This formula provides a basis for a general numerical program
for NNLO QCD computations, combining universal subtraction
ingredients with process-dependent matrix elements.

The subtraction scheme of Ref.~\cite{Devoto:2025jql} relies on a particular parametrization of the phase space of unresolved partons \cite{Czakon:2010td,Czakon:2011ve}. However, it does not constrain the parametrization of the phase space of resolved partons (i.e., the underlying Born kinematics), and a judicious choice of this parametrization can have a dramatic impact on the efficiency of the numerical implementation.

Two further challenges arise in the numerical evaluation.
First, the partitions and sectors needed to disentangle singularities lead to a proliferation of
  contributions that have to be evaluated by Monte Carlo integration. Typically, there are
  significant cancellations between different
  terms, making an efficient organization of their integration challenging. Second, the required Monte Carlo integrations are high-dimensional. For the two physical processes that we consider in this paper, one has to integrate over up to sixteen
  variables.

\section{Implementations of the general formula}
\label{sec:development}
The numerical implementation of the general
formula for finite NNLO QCD remainders~\cite{Devoto:2025jql} for a specified
process requires a large number of ingredients.
In addition
to a proper understanding of this formula, one needs to enumerate partonic channels,
compute or assemble all
double-virtual, real-virtual and double-real amplitudes, design a phase-space parametrization
for the Born process and combine it with the (fixed) parametrization of the unresolved phase space~\cite{Czakon:2010td,Czakon:2011ve}, derive explicit expressions for the action of the soft and collinear limit operators, including terms that depend on the phase-space parametrization, and combine them with the reduced matrix elements. When all the ingredients are in place, one has to write a code that performs the actual numerical integration
and computes kinematic distributions.

Furthermore, one has to perform a series of numerical checks to establish the validity of the code as thoroughly as possible. This includes checking that the counterterms remove non-integrable singularities in all soft, collinear and overlapping limits, that the matrix elements used agree with external providers, that results are independent of technical cuts introduced for Monte Carlo integration and other auxiliary parameters, that the NNLO QCD implementation for the final state $\rmX$ reproduces the NLO QCD
prediction for $\rmX+j$ when one requires an additional resolved jet, and that the same
implementation gives correct NNLO results for simpler but related processes.

These programs are usually written by scientists with knowledge of the underlying
physics and practical experience with similar computations. In fact,
as mentioned in Section~\ref{sec:introduction}, we have
developed such an implementation over the course of a year, using Higgs production in WBF and Higgs production in WBF in association with an additional jet as test cases. This code serves as a benchmark for the codes produced by the LLM agents.

\begin{figure}[t]
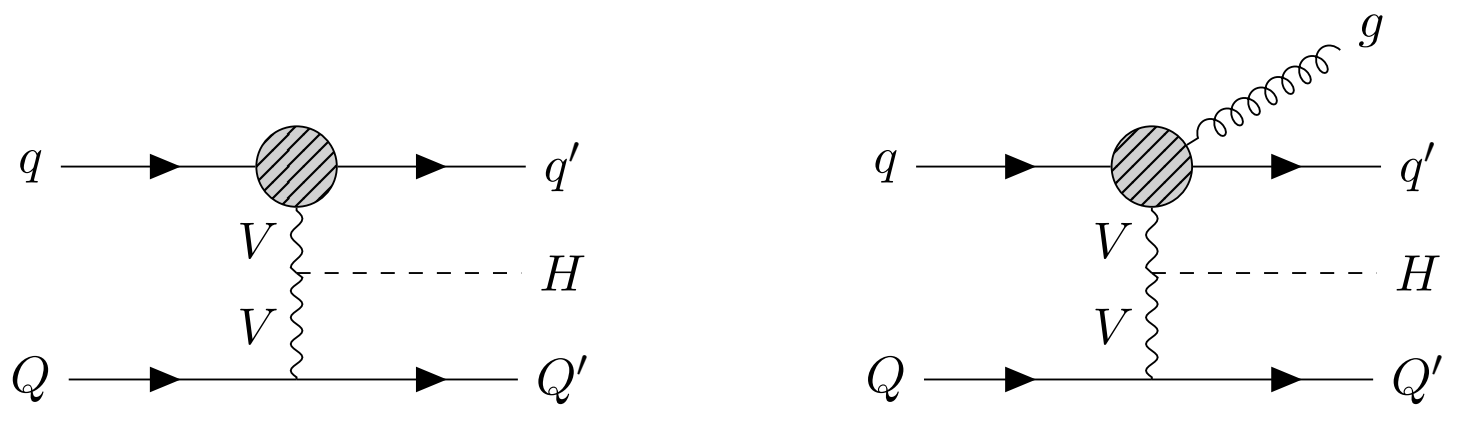

    \centering
    \WBFBornFigure
    \caption{WBF (left) and WBF+$j$ (right), with $V=W,Z$.
    The blob denotes QCD corrections to the upper quark line.
    The lower line is kept at Born level.}
    \label{fig:wbf_born}
\end{figure}

We note that, for both processes, we work in the double deep-inelastic scattering (double-DIS) approximation
and, for the purposes of this paper, we restrict QCD radiation and loop corrections to the upper
fermion line, see Figure \ref{fig:wbf_born}.
This restriction nevertheless retains all the subtraction and
integration challenges needed for the study pursued here.

Using professional subscriptions to Anthropic Claude and OpenAI ChatGPT, we tasked LLM agents with constructing new codes for
computing NNLO QCD corrections for WBF and WBF+$j$, starting from the general formula
of Ref.~\cite{Devoto:2025jql} and related earlier papers~\cite{Devoto:2023rpv, Devoto:2025kin}.  In order to understand to what extent the coding agents require or benefit from human guidance, we pursued two routes with two distinct groups of coding agents.
The first group worked under the close supervision of a human expert, while the second developed the code largely autonomously, with limited human intervention. We use the labels \textsc{Guided LLM} and \textsc{Autonomous LLM} to distinguish these two approaches, the codes that they produced, and the physics outputs of these codes; we emphasize that they refer to development workflows, not to individual agents, models or providers. We supplied the coding agents with the amplitudes used in the human code for the WBF and WBF+$j$ processes~\cite{DeLaurentis:2026dpj,Gehrmann:2005pd,Garland:2002ak,Bern:1997sc}. The double-virtual matrix elements for WBF+$j$ have been constructed starting from the results in Refs~\cite{Garland:2001tf, Garland:2002ak, Gehrmann:2011ab}
available in the public \texttt{NNLOJET} code~\cite{NNLOJET:2025rno}.
Finally, we used the adaptive Monte Carlo algorithm \texttt{VEGAS}~\cite{Lepage:1977sw} as the common integrator and asked both groups to
include histogramming routines for computing distributions of observables.

The first group of agents was tasked with constructing a WBF implementation from scratch with extensive guidance from human experts. We supplied the general formula of Ref.~\cite{Devoto:2025jql}, additional details from Refs~\cite{Devoto:2023rpv,Devoto:2025kin}, optimized amplitude routines and  general analytic expressions for the soft and collinear limits of the finite remainders.
We provided the source code of Ref.~\cite{Asteriadis:2021gpd} and instructed the agents to use it solely for numerical benchmarking. This restriction reflects the fact that the subtraction scheme of Ref.~\cite{Asteriadis:2021gpd}, although closely related to that of Ref.~\cite{Devoto:2025jql}, differs in key technical details.\footnote{For example, the calculation of Ref.~\cite{Asteriadis:2021gpd} employs a slightly simpler, but less general, phase-space partitioning tailored for WBF-like processes.}
Development of the code took place on a laptop without direct cluster access. A human expert submitted jobs to a cluster, provided their output  to the agents, and discussed the results with them, directing the investigation of discrepancies and numerical instabilities and identifying performance bottlenecks for them to address. Once the WBF implementation had been validated, the agents extended it to WBF+$j$ at our request, using the additional amplitudes we supplied.

\begin{table}[!t]
    \centering
    \caption{Inputs and working conditions for the two LLM implementations.
    The row labeled \emph{Additional limits} refers to analytical expressions needed when one emitted parton becomes soft or collinear, beyond those given explicitly in Ref.~\cite{Devoto:2025jql}.}
    \resizebox{\linewidth}{!}{\VBFImplementationTable}
    \label{tab:inputs}
\end{table}

The autonomous route began with the exploratory development of a code to compute $pp \to H+j$ at NNLO in QCD, using the general formula of Ref.~\cite{Devoto:2025jql} with additional details from Refs~\cite{Devoto:2023rpv,Devoto:2025kin}. We also provided the coding agents with amplitude routines for this process and a human-written inclusive Higgs-production program from earlier gluon-fusion studies~\cite{Caola:2019nzf} as an example. Comparisons of the resulting $H+j$ code with \texttt{NNLOJET}~\cite{NNLOJET:2025rno,Chen:2014gva} were successful and encouraged us to extend the exercise to WBF and WBF+$j$, taking the LLM-created $H+j$ code as a starting point. The agents, whom we supplied with the optimized WBF and WBF+$j$ amplitudes used in the human code, derived and implemented the required soft and collinear limits from their implicit definitions in Ref.~\cite{Devoto:2025jql}. Direct cluster access allowed the agents to compile code, submit jobs, inspect output and investigate failures on their own. A human expert defined tasks and supplied information when needed, but between these interventions code development proceeded \emph{largely unsupervised}, including the choice of tests and the organization of the numerical workflow. Once the autonomous agents had constructed and validated their WBF code, they extended it to WBF+$j$.

The two routes thus differ in the degree of human involvement, access to reference source code and the computational resources employed. A summary of the inputs and further details can be found in  Table~\ref{tab:inputs}. Having described the tasks assigned to the two groups of LLM agents, we discuss their approaches and the numerical codes they produced in the following sections.

\section{Remarks}
\label{sec:development_lessons}

In this section, we summarize 
important observations and insights that we obtained from our experiment with LLM coding agents  described above. 

\par\addvspace{\baselineskip}
\noindent\textbf{Fast and reliable implementations are possible:}
\noindent
The main conclusion of our experiment is that coding agents
\emph{can turn the general formula of Ref.~\cite{Devoto:2025jql} into reliable working codes}
for computing NNLO QCD corrections to complex processes, even with relatively
modest human input and guidance. Section~\ref{sec_Results} provides direct evidence for this statement by comparing the results and performance
of the LLM-generated codes with those of the human reference code. Here, we discuss the development time and required resources.

The human reference code took about a year to develop. For each LLM implementation,
generating the initial code required only a few days, while testing, validation
and the calculations discussed in Section~\ref{sec_Results} were completed within a few weeks.
The agents used a few billion tokens per code across development, testing,
validation, and revisions.
The ease and speed of this development open up
exciting opportunities for automating the subtraction aspect of NNLO calculations, which we discuss
further in the conclusions. We recall, however, that the human reference code was
designed with a modular structure that allows its components to be reused
across processes. Therefore, implementing a NNLO QCD computation for a  different collider process would require substantially less time than the initial development of the code.

\par\addvspace{\baselineskip}
\noindent\textbf{Independent validations:}
\noindent
The agents in the autonomous group performed a range of checks to validate their implementation.
For example, one agent identified a
mistake in one of the functions entering the finite remainders in
Ref.~\cite{Devoto:2025jql} by checking the renormalization-scale dependence of a cross section
at NNLO. However, the same group omitted the standard check of ensuring  that the WBF cross section
is numerically independent of the technical cuts used in Monte Carlo integration.
Instead,  the  agents initially adopted the cut values from an earlier implementation for a different process. They then discovered  
discrepancies with a publicly-available code and resolved the problem by investigating the issue of technical cuts. 
This example illustrates the need to scrutinize the results produced by coding agents,
while also demonstrating their usefulness in carrying out validation checks.
By explicitly guiding the agents to perform the checks that human experts routinely employ,
such as those listed at the beginning of Section~\ref{sec:development},
we can make LLM-generated codes \emph{as reliable} as those produced by human experts. 

\par\addvspace{\baselineskip}
\noindent\textbf{Phase-space parametrization:}
\noindent
The guided agents proposed a phase-space parametrization that generates the Higgs rapidity directly from the \texttt{VEGAS} variables, whereas the original parametrization in the human code generated the virtuality of the $t$-channel electroweak gauge bosons~\cite{Asteriadis:2021gpd}. The proposed parametrization proved to be
substantially more efficient: 
for WBF, the gain in integration efficiency at fixed statistical precision ranges from approximately four to more than twenty, depending on the applied cuts and perturbative order, as discussed in Section~\ref{sec_Results}. The agents in the autonomous group also used the Higgs rapidity as an integration variable, together with the transverse momenta and rapidities of the outgoing partons. This parametrization also proved more efficient than the one originally used in the human code.

\par\addvspace{\baselineskip}
\noindent\textbf{Matrix elements:}
\noindent
We supplied the agents with amplitude routines for
WBF and WBF+$j$  optimized by human experts.  We then asked the agents to reduce their evaluation time further.
For the routines used in the real--virtual contribution to WBF+$j$, the LLM agents
achieved a further speedup by a factor of two to three. The agents also made it straightforward
to check our matrix elements against independent implementations by setting up 
and running numerical comparisons with \texttt{MadGraph5\_aMC@NLO}~\cite{Alwall:2014hca}
and \texttt{GoSam}~\cite{Cullen:2011ac}.

\par\addvspace{\baselineskip}
\noindent\textbf{Numerical stability:}
\noindent
In the \textsc{Guided LLM} implementation, substantial effort was also devoted 
to maintaining numerical stability in double precision near soft and
collinear limits, where large contributions must cancel accurately.
Under the guidance of a human expert, the agents analytically rewrote numerically sensitive expressions, e.g.\ limits of matrix elements or splitting functions,  to improve their stability and reduce the loss of precision from rounding errors. Coding agents also made it easier to identify numerically unstable parts of the implementation.
In the \textsc{Autonomous LLM} implementation, the agents likewise identified numerical instabilities by testing cancellations in soft and collinear limits and examining the largest event weights. For example, where direct evaluation of an amplitude lost precision, they extrapolated its value from nearby phase-space points where the evaluation was stable. This allowed them to retain double precision. They also varied the technical cuts used in sensitive regions over several orders of magnitude and found the results stable within statistical uncertainties.

\par\addvspace{\baselineskip}
\noindent\textbf{Run strategies:}
\noindent
Interestingly, it turns out that the integration strategies chosen by coding agents are not always the most
effective, particularly for more demanding calculations. In the WBF+$j$
implementation, the \textsc{Autonomous LLM} agents allocated too few sampling
points to grid training in the multichannel setup, leading to unreliable error estimates and limiting the quality of importance sampling by the training noise. We therefore asked them to revise the run strategy
and devote more resources to training the individual grids. Moreover, when computing transverse-momentum distributions, the agents found that a change of variables that improved the precision of the total cross section produced large fluctuations in some bins. They therefore reversed the change of variables when computing distributions. Under our guidance, both groups of agents also revised how they grouped contributions for integration, combining terms whose sums depend less strongly on the integrated soft and collinear subtraction terms.

\par\addvspace{\baselineskip}
\noindent\textbf{Code optimization:}
\noindent
LLMs proved effective at code optimization, but their first implementations
still contained avoidable overheads: they stored results that were rarely
reused, repeated calculations and evaluated unnecessary quantities.
In both development modes, we made successive requests for improvements,
assessing the resulting performance before asking the agents to refine the
code further. This iterative process was essential to achieving the desired
efficiency.

\section{Results}
\label{sec_Results}
\begin{table}[t!]
    \centering
    \caption{Inclusive and fiducial WBF results from the \textsc{Humans} code
    with old and new phase-space parametrizations, the
    \textsc{Guided LLM} code and the \textsc{Autonomous LLM} code.
    LO entries give cross sections, while NLO and NNLO entries are
    additive corrections, not cumulative predictions.
    Each block lists $\sigma\pm\delta\sigma$ (fb), production time in CPU core-hours
    (referenced to a single core of an AMD EPYC 9755 processor), and the dimensionless ratio
    $\calE/\calE_0$, with $\calE$ defined in \eq\eqref{eq:efficiency}.
    Timings exclude \texttt{VEGAS} grid training and histogram filling.}
    \resizebox{\linewidth}{!}{\VBFComparisonTable}
    \label{table_wbf_comparison}
\end{table}
Following the approach described above, we produced numerical codes for fully differential calculations of WBF and WBF+$j$ production in hadron collisions through NNLO in perturbative QCD.  They include: 
\begin{enumerate}[label=\roman*)] 
    \item a human-written code for WBF and WBF+$j$. 
    The WBF code uses both the original Born phase-space parametrization~\cite{Asteriadis:2021gpd} and the more efficient one proposed by the agents, as discussed in Section~\ref{sec:development_lessons}. For the WBF+$j$ calculation, we use only the latter.
    
    \item  a code for WBF and WBF+$j$  
    produced by a group of agents under close supervision of a human expert (\textsc{Guided LLM});
    
    \item a code for WBF and WBF+$j$ produced by a second group of agents with limited human intervention (\textsc{Autonomous LLM}).
\end{enumerate}

\noindent We remind the reader that we work in the double-DIS approximation and include QCD corrections only on the upper quark line (cf.\ Figure~\ref{fig:wbf_born}). Thus the results presented here are not intended for phenomenological studies, but serve to compare the numerical codes.

Tables~\ref{table_wbf_comparison} and~\ref{table_wbfj_comparison} show the results for WBF and WBF+$j$, respectively. LO entries give cross sections, while NLO and NNLO entries are additive corrections, not cumulative predictions.
The fiducial selection criteria and the physical input parameters are summarized in
Appendix~\ref{app:numerical_setup}.
The tables also give the runtimes for a target Monte Carlo uncertainty of $\delta\sigma=1.0\,\mathrm{fb}$. This amounts to a few parts in ten thousand of the Born cross section for WBF and about two per mille for WBF+$j$. The quoted Monte Carlo integration uncertainties on the NNLO corrections themselves are roughly $1$--$3\%$. The codes were run under similar conditions, with timings referenced to a single core of an AMD EPYC 9755 processor. For a given process, the runtimes at the same statistical precision thus provide a measure of relative efficiency.
\begin{table}[t!]
    \centering
    \caption{Fiducial WBF+$j$ results, with quantities and conventions
    as in Table~\ref{table_wbf_comparison}. Results with the old human
    parametrization are not reported.}
    \resizebox{\linewidth}{!}{\VBFJetComparisonTable}
    \label{table_wbfj_comparison}
\end{table}

The same absolute uncertainty corresponds to different relative precisions for WBF and WBF+$j$. To compare their computational costs at a common precision relative to the Born cross section, we define the \emph{integration difficulty} as
\begin{equation}
    \calE =
    \left(
        \frac{\delta\sigma}{\bar\sigma_{\mathrm{LO}}}
    \right)^{\mathclap{2}} \times T \, ,
    \label{eq:efficiency}
\end{equation}
where $\delta\sigma$ is the statistical uncertainty on the cross section after $T$ CPU core-hours of Monte Carlo integration using pre-trained grids. We use $\bar\sigma_{\mathrm{LO}}$ as the reference Born cross section for each process and selection ($4092\,\mathrm{fb}$ for inclusive WBF, $2776\,\mathrm{fb}$ for fiducial WBF, and $573\,\mathrm{fb}$ for fiducial WBF+$j$).

For a fixed sampler with finite variance and $N$ integration points, let $C$ denote the standard deviation of the sampling weight in units of $\bar\sigma_{\mathrm{LO}}$. We then have
\begin{equation}
    \delta\sigma \simeq
    \frac{C}{\sqrt{N}} \, \bar\sigma_{\mathrm{LO}} \, ,
    \qquad
    T  \simeq N \times \bart \,,
\end{equation}
consequently, 
\begin{equation}
    \calE \simeq C^2 \times \bart \, ,
\end{equation}
where  $\bar t$ is
the mean evaluation time per sampling point. 
Thus, $\calE$ accounts for both the fluctuations in the integration weights and the mean evaluation time per point. Under the assumed scaling, it is approximately independent of the requested precision.
Tables~\ref{table_wbf_comparison} and~\ref{table_wbfj_comparison} report $\calE/\calE_0$, with $\calE_0=10^{-8}\,\mathrm{h}$ corresponding to a relative uncertainty $\delta\sigma/\bar\sigma_{\mathrm{LO}}=10^{-4}$ reached in one CPU core-hour.

The runtimes shown in Tables~\ref{table_wbf_comparison} and~\ref{table_wbfj_comparison} exclude the cost of Monte Carlo grid training, which can be substantial. For the NNLO correction to WBF+$j$, training can take up to a few thousands  CPU core-hours, compared with about $1000$--$2000$ CPU core-hours for production with the LLM-assisted and human-written codes.
We exclude 
these costs for three reasons.
First, we are interested in a comparison of production efficiency at fixed statistical precision with fixed, nearly optimal grids, rather than the total preparation and running cost. Second,  for phenomenological applications,  WBF and WBF+$j$ codes would include pre-trained grids, leaving potential 
users to refine them for their selection criteria rather than perform full training anew. Third, training costs depend on the learning rate, schedule, and stopping criterion, which we neither optimized nor standardized across different implementations.

Beginning with Table~\ref{table_wbf_comparison}, we observe that the inclusive and fiducial cross sections and their higher-order corrections in all four columns are statistically consistent. This agreement confirms the conclusion of the previous section that LLM coding agents can implement the subtraction formulas of Ref.~\cite{Devoto:2025jql} in a working numerical code. Looking at the run times for the human-written reference code, we see the substantial improvement in speed achieved by using the new parametrization discussed in the previous section.  
At NNLO, the two LLM codes and the human code with the new parametrization have comparable runtimes, with the guided code fastest for both selections. The human code with the old parametrization takes substantially longer.
\begin{figure}[t]
    \centering
    \includegraphics[width=0.49\linewidth]{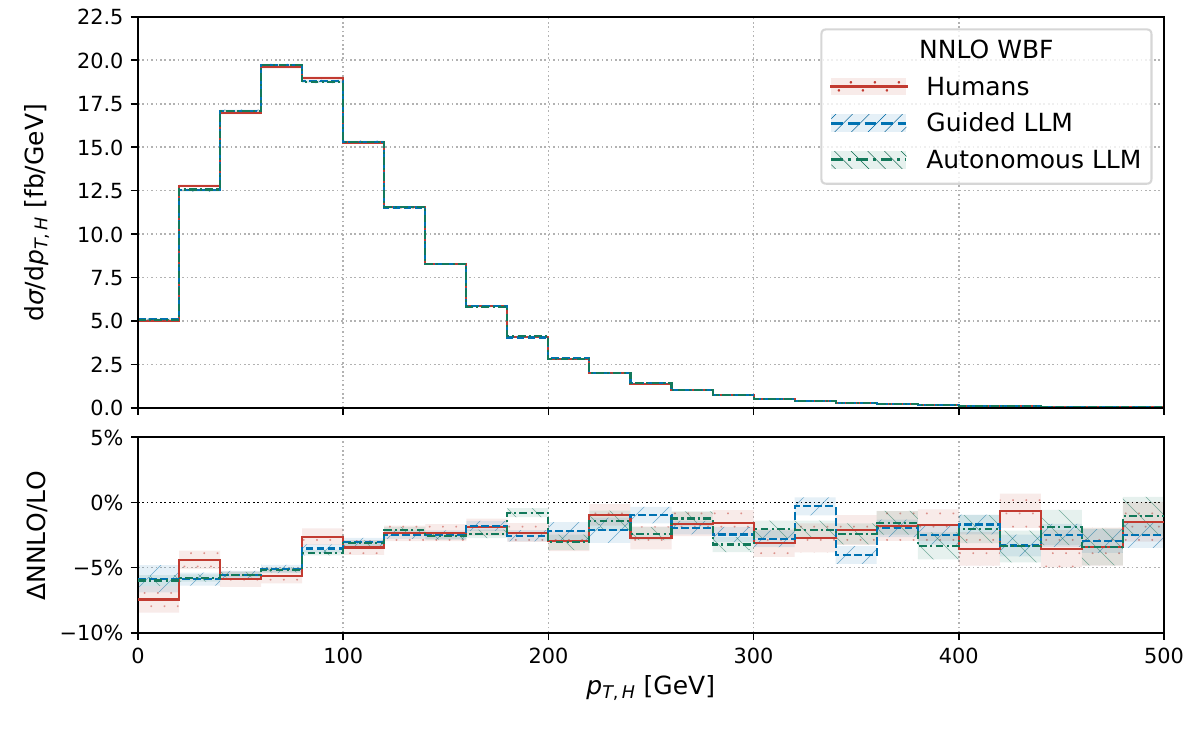}
    \includegraphics[width=0.49\linewidth]{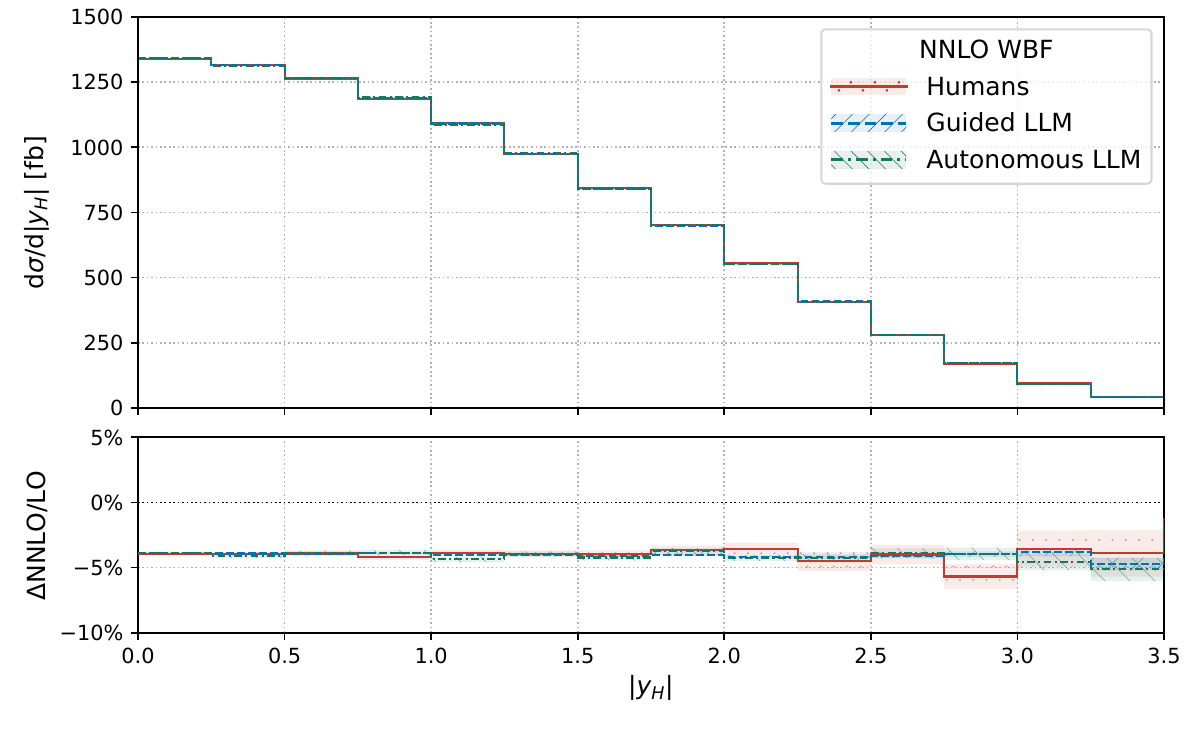}\\
    \includegraphics[width=0.49\linewidth]{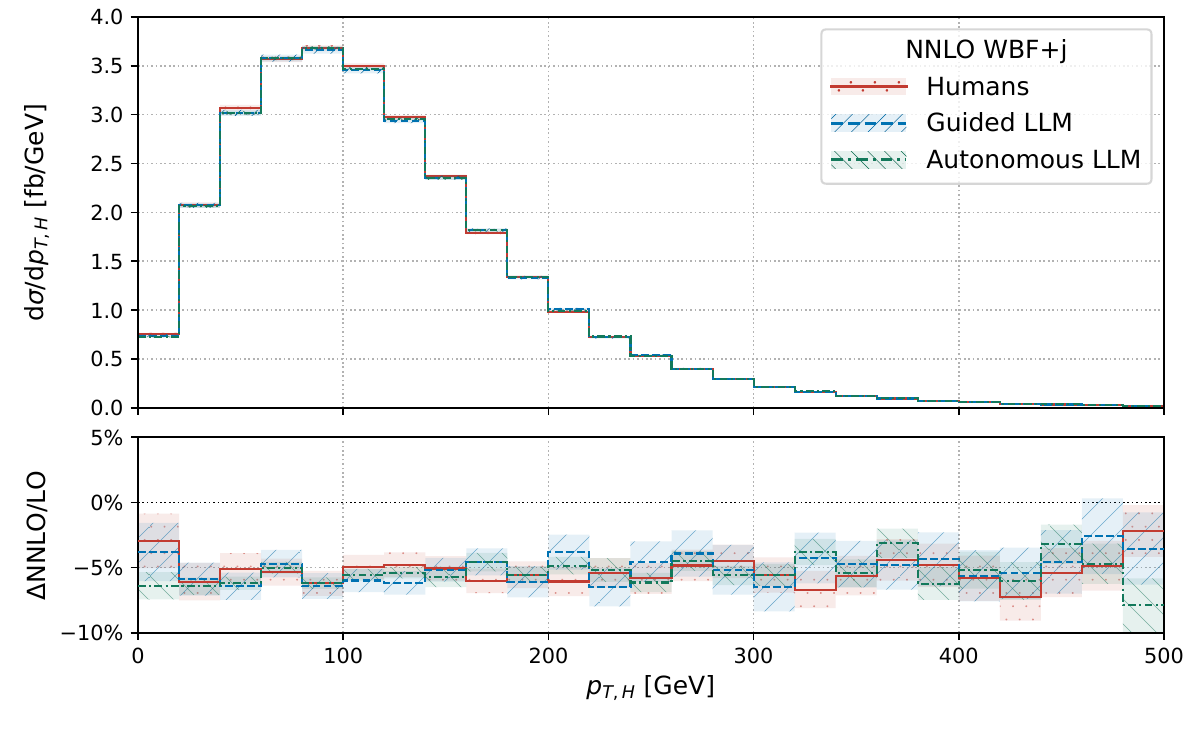}
    \includegraphics[width=0.49\linewidth]{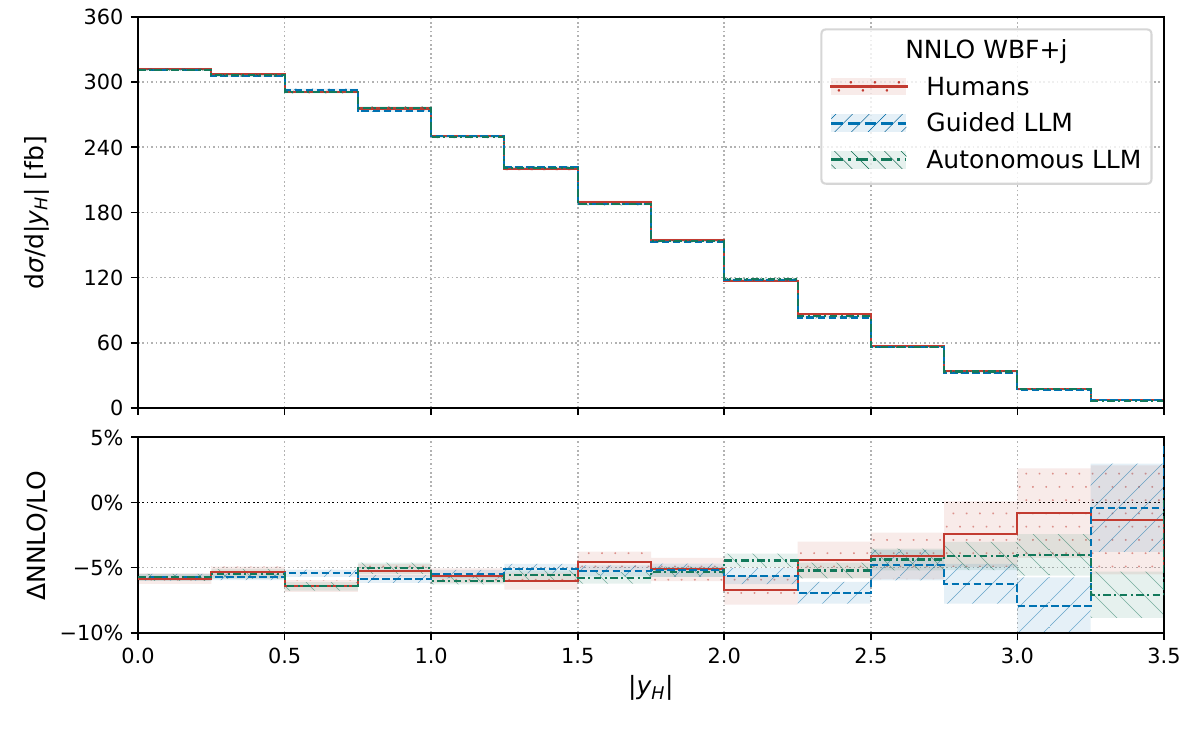}
    \caption{Higgs-boson transverse-momentum and absolute-rapidity distributions in fiducial WBF (upper plots) and WBF+$j$ (lower plots), comparing the \textsc{Humans} code with the  new phase-space parametrization (solid red), \textsc{Guided LLM} (dashed blue) and \textsc{Autonomous LLM} (dash-dotted green) implementations. The upper panel of each plot shows the differential cross section including both the NLO and NNLO corrections. The lower panels show the additive NNLO correction in each bin as a percentage of the corresponding LO prediction.}
    \label{fig:human_vs_codex_distributions}
\end{figure}

Turning to the results for WBF+$j$ reported in Table~\ref{table_wbfj_comparison}, 
we again observe agreement within the statistical uncertainties between the codes for the LO cross section and the NLO and NNLO corrections. Compared with fiducial WBF, the increase in integration difficulty grows strongly with perturbative order; for this reason, 
we ran the human code only with the new phase-space parametrization.
At NNLO,  all codes have similar runtimes and $\calE$ values.

Finally, we show the distributions in Higgs transverse momentum and absolute rapidity for the WBF and WBF+$j$ processes in Figure~\ref{fig:human_vs_codex_distributions}. The three codes also agree for both differential distributions. Obtaining the statistical precision shown in Figure~\ref{fig:human_vs_codex_distributions} required roughly three times the production time quoted in Tables~\ref{table_wbf_comparison} and~\ref{table_wbfj_comparison}.{\interfootnotelinepenalty=10000\footnote{The method of moments in Ref.~\cite{MelnikovNovikovPedron2026} could reduce this extra runtime.}}

\section{Conclusions}
\label{sec_Conclusions}

In this paper, we have shown that  LLM coding agents can provide reliable 
numerical implementations of NNLO QCD corrections
to specified processes, starting from the analytic   
formula  for infrared subtractions given in   Ref.~\cite{Devoto:2025jql}. Although this 
formula is complicated and partially symbolic, 
this did not present an obstacle for coding 
agents, who were able to implement 
it in numerical codes 
with varying degrees of human guidance. The modular structure of the formula allowed both the closely guided and  largely autonomous groups of coding agents to extend their WBF codes to WBF+$j$ by reusing the same subtraction machinery with the additional amplitudes for that process.

We have used two processes -- Higgs production in 
WBF and  Higgs production in association with a jet in WBF -- to study the quality of LLM-assisted implementations.  Computing NNLO QCD corrections to such  processes 
is known to be complicated because they have many particles already at leading order, strongly forward kinematics, a complicated flavor structure, and small  radiative corrections. As such, they provide an excellent testing ground to showcase the impressive capabilities of the coding agents. 

Some of the most striking aspects of their performance are 
{\it i)}~dramatically 
reduced time needed to implement  NNLO QCD corrections 
for a new process provided that relevant amplitudes are available; 
{\it ii)}~a low cost of testing new ideas and {\it iii)}~the  
ease with which various checks  that ensure the reliability of the calculation can be performed.

Establishing the reliability of LLM-assisted numerical codes is an important part of implementing NNLO QCD corrections for a new process. This can be done by a standard set of validation tests which can be performed expeditiously by the coding agents, although expert (human) judgment is still beneficial in guiding these tests and directing investigations into any discrepancies that might arise. Once an LLM-written code passes these tests, it can be considered  as  robust as a human-written code that passes the same checks.

Given our findings, one can speculate about  broader implications of the fact that coding agents are able to implement NNLO QCD corrections for arbitrary processes using the formula in Ref.~\cite{Devoto:2025jql} as a starting point.  Since such implementations seem to require a few days to a few weeks, doing this is very inexpensive.
As a result, it becomes tempting  to  imagine  
a shared, publicly accessible  repository of similarly structured, interconnected numerical 
codes for NNLO QCD description of different processes, 
created in an LLM-assisted 
way.  
With common interfaces and reproducible validation tests, contributions from different actors could be continuously added to the repository.  Thanks to coding agents, new ideas and improvements to shared components can be immediately implemented. 
Such a framework would turn the generality of the formula derived in Ref.~\cite{Devoto:2025jql} into a lasting computational resource for the community, making the extension of NNLO predictions to new processes a cumulative undertaking rather than a succession of separate, private implementations.

\section*{Acknowledgements}
R.R.\ is grateful to the Institute for
Theoretical Particle Physics at KIT for hospitality extended to him during the course of the work on this paper.
I.N. acknowledges the CERN Theoretical Physics Department for its hospitality during the course of this research.
The research of K.M., I.N., and D.M.T. is supported by the Deutsche Forschungsgemeinschaft (DFG, German Research Foundation) under grant no.\ 396021762 - TRR 257.
The research of M.T. is supported by ``Karlsruhe School of Elementary and Astroparticle Physics'' (KSETA) through a grant from ``Deutscher Akademischer Austauschdienst'' (DAAD).
F.D. is supported by the United States Department of Energy, Contract DE-AC02-76SF00515.

\appendix
\section{Numerical setup}
\label{app:numerical_setup}

This appendix specifies the input parameters and event selections used for the numerical comparisons in this paper.
We consider proton--proton collisions at $\sqrt{s}=13\,\mathrm{TeV}$.
The electroweak inputs are
\begin{equation}
\begin{aligned}
    m_H&=125\,\mathrm{GeV}\,, &\qquad G_{\rmF}&=1.16639\times10^{-5}\,\mathrm{GeV}^{-2}\,,\\
    m_Z&=91.1876\,\mathrm{GeV}\,, &\Gamma_Z&=2.4952\,\mathrm{GeV}\,,\\
    m_W&=80.398\,\mathrm{GeV}\,, &\Gamma_W&=2.1054\,\mathrm{GeV}\,,
\end{aligned}
\end{equation}
and the CKM matrix is the unit matrix.
The Higgs boson is treated as stable.

The PDFs and strong coupling $\alpha_{\rms}$ are obtained through the \texttt{LHAPDF}  library~\cite{Buckley:2014ana}.
The same \texttt{NNPDF31\_nnlo\_as\_0118} parton distributions~\cite{Ball:2017nwa} are used at every perturbative order.
We use fixed renormalization and factorization scales,
\begin{equation}
 \mu_{\rmR}=\mu_{\rmF}=125\,\mathrm{GeV}\,.
\end{equation}

The fiducial selection requires at least two jets for WBF and at least 3 jets for WBF+$j$ with transverse momentum $p_{\rmT,j}\geq25\,\mathrm{GeV}$.
The jets are reconstructed with the inclusive anti-$k_{\rmT}$ algorithm~\cite{Cacciari:2008gp}, with the jet radius parameter $R=0.4$.
No constraints are imposed on the rapidity and the invariant mass of the jets.
For inclusive WBF results, all events are accepted and no jet-selection cuts are applied.


\bibliography{biblio}

\end{document}